\documentclass[useAMS,usenatbib]{mn2e}              
\usepackage{epsfig}
\newcommand{\etal}{{et al.~}}

\newcommand{\kms}{\>{\rm km}\,{\rm s}^{-1}}

\newcommand{\beq}{\begin{equation}}
\newcommand{\eeq}{\end{equation}}

\newcommand{\apj}{ApJ}

\newcommand{\apjs}{ApJS}
\newcommand{\aj}{AJ}
\newcommand{\mnras}{MNRAS}

\newdimen\hssize
\newdimen\hdsize 
\title[Shear-galaxy property relation] 
{Connecting the physical properties of galaxies with the overdensity
and tidal shear of the large-scale environment}

\author[J. Lee \& C. Li]
{Jounghun Lee$^{1}$\thanks{E-mail:jounghun@astro.snu.ac.kr}, 
and Cheng Li$^{2,3}$\\
$^{1}$Department of Physics and Astronomy, FPRD, Seoul National
University, Seoul 151-747, Korea \\
$^{2}$Max-Planck-Institute for Astrophysics,
Karl-Schwarzschild-Str. 1, D-85741 Garching, Germany \\
$^{3}$MPA/SHAO Joint Center for Astrophysical Cosmology at Shanghai 
Astronomical Observatory, Nandan Road 80, Shanghai 200030, China
}
\begin{document}

\date{Accepted 2008 ???. Received 2008 ???; in
original form 2008 March 1}

\pagerange{\pageref{firstpage}--\pageref{lastpage}} \pubyear{2008}

\maketitle

\label{firstpage}

\begin{abstract}

We have examined the  correlations between the large-scale environment
of galaxies  and their physical  properties, using a sample  of 28,354
nearby  galaxies drawn  from the  Sloan  Digital Sky  Survey, and  the
large-scale  tidal field reconstructed  in real  space from  the 2Mass
Redshift Survey and smoothed over a radius of $\sim 6 h^{-1}$Mpc.  The
large-scale environment is expressed  in terms of the overdensity, the
ellipticity of  the shear and  the type of the  large-scale structure.
The physical  properties analyzed include  $r$-band absolute magnitude
$M_{^{0.1}r}$,  stellar  mass  $M_\ast$, $g-r$  colour,  concentration
parameter $R_{90}/R_{50}$ and surface stellar mass density $\mu_\ast$.

Both luminosity and stellar mass  are found to be statistically linked
to the  large-scale environment, regardless of how  the environment is
quantified. More luminous  (massive) galaxies reside preferentially in
the regions  with higher densities, lower  ellipticities and halo-like
structures. At  fixed luminosity, the  large-scale overdensity depends
strongly on  parameters related to the recent  star formation history,
that  is colour  and  $D(4000)$,  but is  almost  independent of  the
structural parameters $R_{90}/R_{50}$ and $\mu_\ast$. All the physical
properties are  statistically linked to  the shear of  the large-scale
environment  even when  the large-scale  density is  constrained  to a
narrow  range.  This  statistical  link  has been  found  to  be  most
significant in the quasi-linear  regions where the large-scale density
approximates to an order of unity, but no longer significant in highly
nonlinear regimes with $\delta_{\rm LS}\gg 1$.

Our   results  suggest   that   the  initial   conditions  have   made
nonnegligible   contributions   to   establishing  the   environmental
dependence of galaxy  properties. It is expected that  our results may
give  a   new  clue   to  the  unknown   physical  mechanism   of  the
galaxy-environment relationships.
\end{abstract}


\section{Introduction}

The spatial distribution  of galaxies as a function  of their physical
properties   provides  important  constrains   on  models   of  galaxy
formation. Our  understanding of such distribution  has come primarily
from large redshift surveys of  nearby galaxies, for example the Sloan
Digital Sky Survey \citep[SDSS;][]{York-00}.  These surveys have shown
that galaxies  are not  distributed homogeneously, but  in filamentary
structures  that  surround  large   empty  regions,  or  voids.   More
importantly,  galaxies  of  different   properties  are  found  to  be
associated  with different  environments.   Indeed, galaxy  properties
such  as  morphology,  luminosity,  colour,  surface  brightness,  gas
content, mean  stellar age, star formation rate,  and nuclear activity
are  all correlated  with the  overdensity of  the  galaxy environment
\citep{Oemler-74,    Dressler-80,    Postman-Geller-84,   Whitmore-93,
Balogh-01,   Lewis-02,   Martinez-02,   Gomez-03,  Goto-03,   Hogg-03,
Kauffmann-04, Blanton-05, Rojas-05,  Park-07}.  Galaxy colour is found
to  be  the galaxy  property  most  predictive  of the  local  density
\citep{Kauffmann-04, Blanton-05}, while  colour and luminosity jointly
comprise the most predictive pair of properties \citep{Blanton-05}.

Studies  of   galaxy  clustering,  usually   quantified  by  two-point
correlation  function  (2PCF), have  also  found different  clustering
amplitudes  for different  types of  galaxies 
\citep[e.g.][]{Davis-Geller-76, Mo-McGaugh-Bothun-94, Park-94, 
Norberg-01, Norberg-02, Budavari-03, Madgwick-03, Zehavi-02, Zehavi-05,
Li-06}.  These studies  have revealed that
the  galaxies  with higher  luminosities,  higher  stellar masses, red
colours, bulge-dominated morphologies and spectral types indicative of
old   stellar   populations  reside   preferentially   in  the   dense
regions.  

Although the  physical mechanism  for the environmental  dependence of
galaxy properties has yet been fully understood, numerical simulations
and (semi-)analytical work have led us to bring up a standard picture,
according  to  which the  nonlinear  processes  on  small scales  like
hierarchical  merging,  galaxy  interactions,  and tidal  forces  have
nurtured the  environment-galaxy property relations.   However, it has
been realized  from recent observational studies  that the large-scale
distribution of  galaxies may be  also correlated with  their physical
properties.   For instance,  \citet{Li-06} found  that, at  {\em
fixed}  stellar  mass, the  dependence  of  clustering  on colour  and
4000-\AA\ break strength --- measures  of mean stellar age --- extends
over very large  physical scales up to 10 Mpc or  more. This is rather
surprising, because  these scales are significantly  larger than those
of individual  dark matter haloes over which  different galaxies could
have exerted any influence on  each other on a Hubble timescale.  What
this  implies is  that,  the star  formation  history of  a galaxy  is
somehow imprinted at  birth. It is then natural  to speculate that the
role  of  the  initial  cosmological conditions  in  establishing  the
environmental variations  of galaxy properties  is a missing  piece in
the standard picture \citep[see also,][]{Tanaka-04}.

In  order to  find clear  observational  evidence in  support of  this
hypothesis, it is necessary to extend the previous studies not only by
going to large scales, but  also by quantifying the galaxy environment
in a  more explicit manner.   In previous studies, the  environment is
usually expressed in terms of  overdensities in galaxy number that are
estimated in  a fixed sphere/aperture  centered on the  galaxies being
studied.  The  radius of the sphere/aperture  is chosen to  be $\la$ a
few Mpc, as to probe the physical processes occurring inside individual
dark matter haloes.   Compared to the local density,  the 2PCF is more
powerful  in the  sense  that it  encapsulates  information about  how
galaxy properties depend on  the overdensity of the galaxy environment
over a wide range of  physical scales. However, it doesn't account for
the other  aspects of  the large-scale environment,  such as  the {\em
tidal shear}. This is  important given the filamentary distribution of
galaxies on large  scales, which is in fact  a nonlinear manifestation
of the primordial tidal field \citep{Bond-Kofman-Pogosyan-96}.

The  morphology-density  relation  is  one  of  the  most  fundamental
correlations   between   the  properties   galaxies   and  the   local
environment.    Recently,   \citet{Lee-Lee-08}   have   extended   the
quantification of  this relation  to large scales,  using a  sample of
15,882  nearby  galaxies  from  the  Tully Catalogue  that  have  been
well-determined in  morphological type, and the  overdensity and tidal
shear of  the real-space linear  density field reconstructed  from the
2Mass Redshift  Survey (2MRS) and smoothed  with a wiener  filter on a
radius of  $\approx 6  h^{-1}$Mpc. The authors  found that  the galaxy
morphological type  is a strong  function of not only  the overdensity
but also the tidal shear of the large-scale environment.

As   pointed  out   by  \citet{Li-06},   the   standard  morphological
classification scheme mixes elements that depend on the structure of a
galaxy with elements related to its recent star formation history, and
so  it  is by  no  means  that these  two  elements  should depend  on
environment in  the same way.  In  this paper, we extend  the study of
\citet{Lee-Lee-08} by using  a sample of galaxies drawn  from the SDSS
and investigating  the dependence  of the large-scale  overdensity and
tidal  shear on  a variety  of physical  properties,  including colour
($g-r$),   4000-\AA\   break   strength  ($D(4000)$),   concentration
parameter   ($R_{90}/R_{50}$)  and   stellar   surface  mass   density
($\mu_\ast$). The first two quantities, that is, $g-r$ and $D(4000)$,
are parameters  associated with the  recent star formation  history of
the galaxy, whereas the other  two are related to galaxy structure. We
also probe the  dependence on luminosity and stellar  mass.  In future
studies, we plan to examine with more physical properties such as star
formation rate and nuclear activity.

The structure  of this  paper is  as follows.  In  Section~2 we  give a
brief overview  of the observational  data, and explain how  to define
the  shear of  large-scale environment  at the  positions of  the SDSS
galaxies. We  present the observed  links of the galaxy  properties to
the  density  and to  the  shear  of  the large-scale  environment  in
Sections~3 and  4, respectively. In  Section~5 we show how  the galaxy
properties vary with the types  of the large-scale structures in which
they are found. In Section~6 we summarize the results and draw a final
conclusion.

We   assume  a   cosmological   model  with   the  density   parameter
$\Omega_0=0.3$  and  the  cosmological constant  $\Lambda_0=0.7$.   To
avoid the $-5\log_{10}h$ factor, the Hubble's constant $h=1$, in units
of  $100\kms{\rm Mpc}^{-1}$,  is  assumed throughout  this paper  when
computing absolute magnitudes.

\section{Data}

\subsection{The SDSS galaxies and physical quantities}

The galaxy sample analyzed in this study is constructed from the SDSS.
The survey goals are to obtain photometry of a quarter
of  the sky and  spectra of  nearly one  million objects.   Imaging is
obtained    in     the    {\em    u,    g,    r,     i,    z}    bands
\citep{Fukugita-96,Smith-02,Ivezic-04}  with a  special  purpose drift
scan camera  \citep{Gunn-98} mounted  on the SDSS  2.5~meter telescope
\citep{Gunn-06}  at Apache  Point Observatory.   The imaging  data are
photometrically    \citep{Hogg-01,Tucker-06}    and    astrometrically
\citep{Pier-03} calibrated,  and used to  select spectroscopic targets
for the main galaxy sample \citep{Strauss-02}, the luminous red galaxy
sample     \citep{Eisenstein-01},     and     the    quasar     sample
\citep{Richards-02}.  Spectroscopic fibres are assigned to the targets
using an efficient tiling  algorithm designed to optimise completeness
\citep{Blanton-03-tiling}.  The details of  the survey strategy can be
found in  \citet{York-00} and  an overview of  the data  pipelines and
products   is    provided   in   the   Early    Data   Release   paper
\citep{Stoughton-02}. More  details on the photometric  pipeline can be
found in \citet{Lupton-01}.

Our parent sample for this  study consists of 397,344 objects that are
spectroscopically  classified  as  galaxies  and  have  data  publicly
available      in      the       SDSS      Data      Release      Four
\citep[DR4;][]{Adelman-McCarthy-06}.   These  galaxies have  Petrosian
$r$-band magnitudes  in the range $14.5<r<17.77$  after correction for
foreground   galactic   extinction   using   the  reddening   maps   of
\citet{Schlegel-Finkbeiner-Davis-98} and  have redshifts in  the range
$0.005\la z<0.30$,  with a  median $z$  of 0.10.  In  order to  have a
similar  redshift  distribution as  the  2MRS,  we  have selected  the
galaxies  with $z\le  0.04$.  We  also  restrict the  galaxies to  the
apparent magnitude  range $14.5<r<17.6$, as to yield  a uniform galaxy
sample that is complete over the entire area of the survey, as well as
to   the  absolute   magnitude  range   $-23<M_{^{0.1}r}<-17$.   Here,
$M_{^{0.1}r}$  is the  $r$-band  absolute magnitude  corrected to  its
$z=0.1$     value     using     the     $K$-correction     code     of
\cite{Blanton-03-Kcorrection}  and the  luminosity evolution  model of
\cite{Blanton-03-LF}.  The resulting sample includes a total of 28,354
galaxies with a median redshift of $<z>\approx 0.02$.

The galaxies are  then divided into a variety  of different subsamples
according to  their physical parameters,  including absolute magnitude
($M_{^{0.1}r}$),  stellar  mass  ($M_\ast$),  colour  [$(g-r)_{0.1}$],
4000-\AA\ break strength ($D(4000)$), concentration ($R_{90}/R_{50}$)
and stellar surface mass density ($\mu_\ast$).  Here, $(g-r)_{0.1}$ is
the   $g-r$  colour  corrected   to  its   $z=0.1$  value   using  the
$K$-correction  code of \cite{Blanton-03-Kcorrection}.   $D(4000)$ is
the narrow version of the index defined in \citet{Balogh-99}. $R_{90}$
and $R_{50}$ are the radii enclosing  90 and 50 per cent of the galaxy
light  in  the $r$  band  \citep[see][]{Stoughton-02}. The  half-light
radius in the $z$ band and the stellar mass yield the effective stellar
surface mass  density ($\mu_\ast=M_\ast/2\pi r^2_{50,z}$,  in units of
$h^2M_\odot  {\rm kpc}^{-2}$).   The  stellar masses  of galaxies  are
estimated  using their  spectra, and  they are  publicly  available at
http://www.mpa-garching.mpg.de/SDSS.    The  reader  is   referred  to
\citet{Kauffmann-03} and \citet{Brinchmann-04} for details.

\subsection{The large-scale density and shear field}

Now that  the galaxy sample is  constructed, we would  like to measure
the   large-scale    shear   at   the   locations    of   the   sample
galaxies. \citet{Erdogdu-06} has  reconstructed the real space density
and velocity fields smoothed on  the scale of $\approx 6h^{-1}$Mpc, in
a  cubic consisting  of  $64^{3}$  pixel points  with  linear size  of
$400h^{-1}$Mpc, by applying the  wiener filter technique to the galaxy
catalogs     from    the     fully-sky    2Mass     Redshift    Survey
(2MRS). \citet{Lee-Erdogdu-07} have  used this real-space 2MRS density
field to reconstruct the  tidal shear field. Basically, they performed
the   Fourier   transformation  of   the   real-space  density   field
$\delta({\bf  x})$  calculated  the   Fourier  space  tidal  field  as
$T_{ij}({\bf  k})  = k_{i}k_{j}\delta  ({\bf  k})/k^{2}$.  Then,  they
perform the inverse Fourier  transformation of the Fourier space tidal
field to derive the real space tidal field, $T_{ij}({\bf x})$.

By  applying  the  cloud-in-cell   method  to  the  2MRS  tidal  field
reconstructed by \citet{Lee-Erdogdu-07},  we calculate the tidal shear
tensor at the position of  each selected SDSS galaxy. We determine the
three dimensional  positions of the  selected galaxies from  the given
information  on $z$,  $\delta$ and  $\alpha$, assuming  a $\Lambda$CDM
cosmology of $\Omega_{m}=0.3$, $\Omega_{\Lambda}=0.7$ and $h=1$. After
measuring the tidal tensor at  each galaxy position, we diagonalize it
to           find           its           three           eigenvalues,
$\{\lambda_{1},\lambda_{2},\lambda_{3}\}$,                       (with
$lambda_{1}\ge\lambda_{2}\ge\lambda_{3}$).   The  trace  of the  tidal
tensor at  the position  of a given  galaxy equals the  linear density
contrast of  the large-scale environment ($\delta_{\rm  LS}$) in which
the given galaxy is embedded:
\begin{equation}
\label{eqn:dls}
\delta_{\rm LS} \equiv \lambda_{1} + \lambda_{2} + \lambda_{3}.
\end{equation} The shear of the  large scale environment is defined as
the  ellipticity   of  the  gravitational   potential  ($e_{\rm  LS}$)
\citep{bar-etal86} as
\begin{equation}
\label{eqn:els}
e_{\rm LS} \equiv \frac{\lambda_{1}-\lambda_{3}}{2\delta_{\rm LS}}.
\end{equation}  We  calculate the  large  scale density  ($\delta_{\rm
LS}$) and shear  ($e_{\rm LS}$) at the positions  of the selected SDSS
galaxies by eqs.~[\ref{eqn:dls}] and [\ref{eqn:els}].

\section{Correlation  of  the   large-scale  overdensity  with  galaxy
properties}

In  this section  we  study  how the  overdensity  of the  large-scale
environment of galaxies depends on their physical properties.

\begin{figure}
\begin{center}
\includegraphics[width=0.5\textwidth,clip=true]{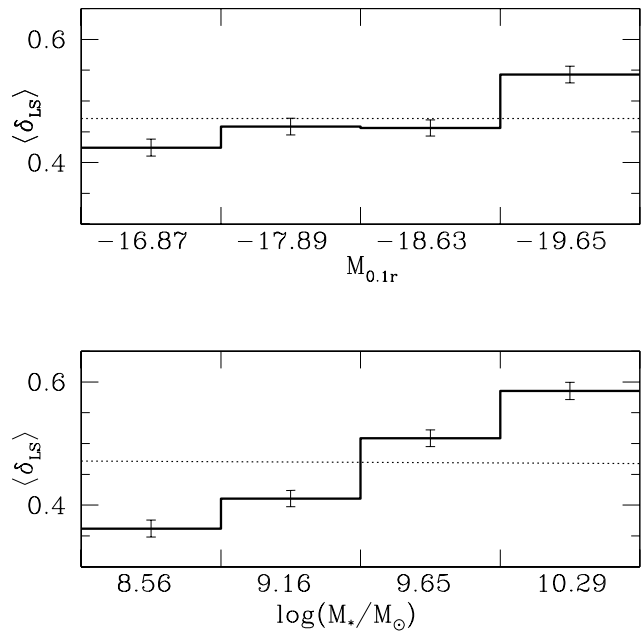}
\end{center}
\caption{Mean of the large scale density ($\delta_{\rm LS}$)
averaged over four different samples of the SDSS galaxies divided by 
their r-band absolute magnitude (upper) and stellar mass (lower).}
\label{fig:den1}
\end{figure}

We first divide  the SDSS galaxy sample described  in \S~2.1 into four
subsamples according  to absolute magnitude $M_{^{0.1}r}$,  as well as
four  subsamples  according  to stellar  mass  $\log(M_\ast/M_\odot)$.
Table~\ref{tab:sub1}   lists   the   ranges   of   $M_{^{0.1}r}$   and
$\log(M_\ast/M_\odot)$ for these subsamples,  which are chosen in such
a  way  that each  subsample  has  approximately  the same  number  of
galaxies.  For  each subsample,  we then calculate  the mean  value of
$\delta_{\rm LS}$, the overdensity of the large-scale environment (see
\S~2.2). Fig.~\ref{fig:den1} plots $\delta_{\rm LS}$ as histograms, as
function of  the mean absolute  magnitude (upper panel)  and stellar
mass (lower  panel) of the  subsamples.  The errors are  calculated as
$\sigma/\sqrt{N_{g}}$  where  $\sigma$ is  one  standard deviation  of
$\delta_{\rm LS}$ and $N_{g}$ is  the number of the galaxies belonging
to  a given  subsample.   The average  result  of the  whole sample  is
plotted as a dotted line in both panels.

As can  be seen from Fig.~\ref{fig:den1}, both  luminosity and stellar
mass are statistically linked to the large-scale density, that is, the
mean value of  the overdensity, $\langle\delta_{\rm LS}\rangle$, increases
with  increasing luminosity and  stellar mass.   It is  interesting to
notice  that  the  trend  of $\langle\delta_{\rm  LS}\rangle$  with  $\log
(M_{*}/M_{\odot})$ is  stronger than  the trend with  $M_{0.1r}$. This
suggests that stellar  mass is more susceptible to  the effect of the
density of large-scale environment.

\begin{figure}
\begin{center}
\includegraphics[width=0.5\textwidth,clip=true]{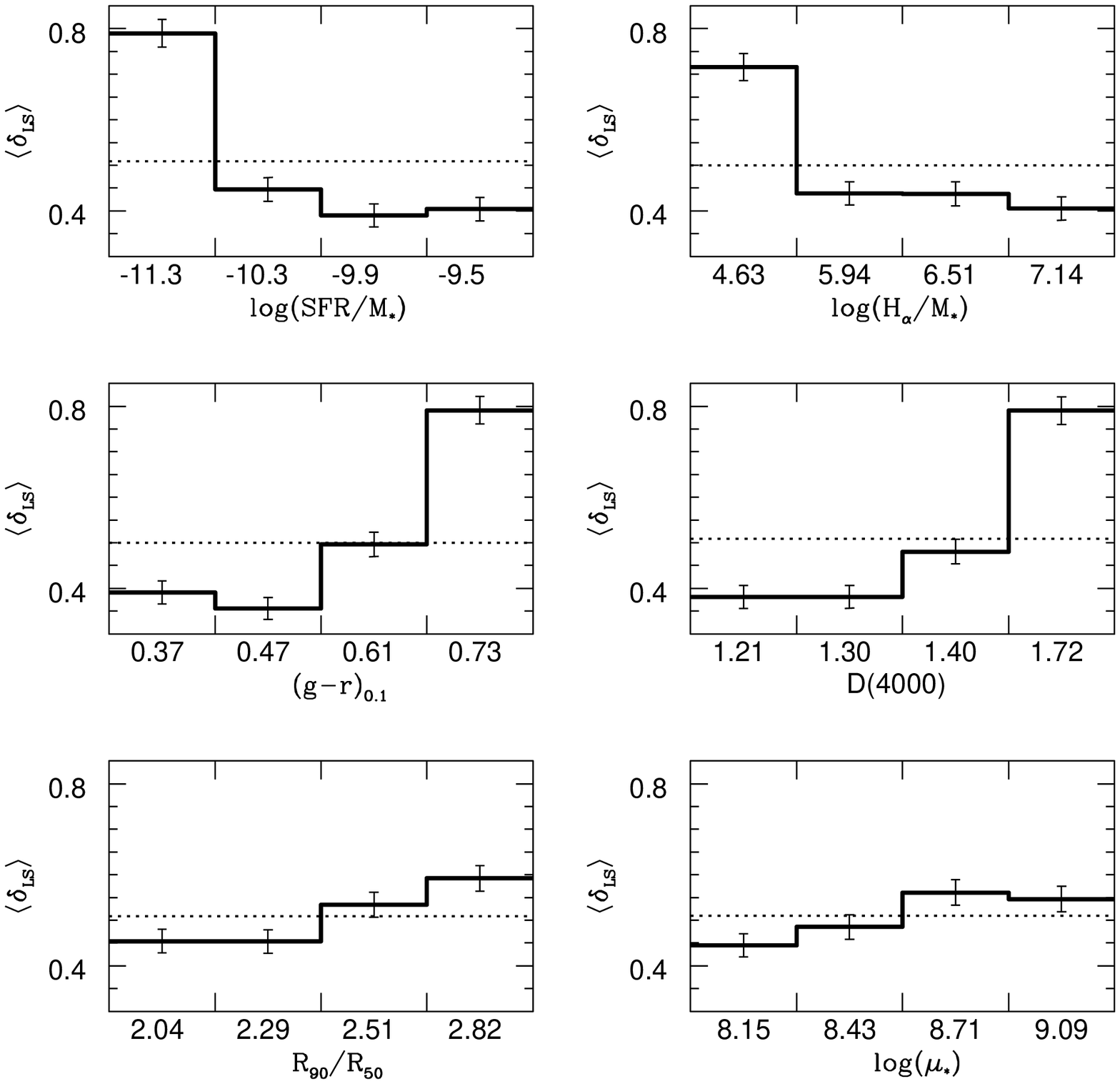}
\end{center}
\caption{Mean of the large scale density ($\delta_{\rm LS}$)
averaged over four different samples of the SDSS galaxies divided by 
their $g-r$ color (top-left), 4000-\AA\ break strength (top-right), 
concentration parameter (bottom-left) and surface stellar mass
density (bottom-right) from SDSS DR4.\label{fig:den2}}
\end{figure}

Next, we  study the dependence  of $\langle\delta_{\rm LS}\rangle$  on the
other    four   physical   quantities:    $(g-r)_{0.1}$,   $D(4000)$,
$R_{90}/R_{50}$  and  $\mu_\ast$.   Since  all  these  quantities  are
correlated  with   galaxy  luminosity   and  stellar  mass,   we  have
restricted this  analysis to  the third luminosity  subsample (Sample
III) in Table~1. The  results are plotted in Fig.~\ref{fig:den2}. Each
panel corresponds to one physical quantity, according to which all the
galaxies in Sample III are divided further into four subsamples in the
same way as described above.  Details of these subsamples are given in
Table~\ref{tab:sub2}.

It  is   seen  that  the   large-scale  overdensity  $\langle\delta_{\rm
LS}\rangle$  depends  strongly  on  parameters related  to  recent  star
formation  history, that  is, $(g-r)_{0.1}$  and $D(4000)$,  but very
weekly     on     structural     parameters    $R_{90}/R_{50}$     and
$\mu_\ast$. ``Old'' galaxies with  redder colours and higher values of
$D(4000)$ are  preferentially located in  the large-scale environment
of higher densities, with  the colour ($D(4000)$) dependence becoming
more   remarkable   for  the   galaxies   with  $(g-r)_{0.1}\ga   0.7$
($D(4000)\ga  1.5$).  These  results   are  in  good  agreement  with
\citet{Li-06}.  We  would like  to point out  that the week  trends of
$\langle\delta_{\rm LS}\rangle$  with $R_{90}/R_{50}$ and  $\mu_\ast$ seen
from the bottom panels of Fig.~\ref{fig:den2} probably follow from the
trends   with    luminosity   and   stellar   mass    as   seen   from
Fig.~\ref{fig:den1},  since  the luminosity  range  of  Sample III  in
Table~1 may be not narrow enough.  We would come back to this point in
future with larger samples.

\begin{table}
\centering
\caption{The range of the r-band absolute magnitude, the stellar mass,
the specific star formation rate, and the $H_{\alpha}$-emission line
of the SDSS galaxies belonging to each subsample \label{tab:sub1}}
\begin{tabular}{@{}ccc@{}}
\hline
Sample & $M_{0.1r}$ & $\log [M_{*}/M_{\odot}]$ \\ \hline
 I  & $> -17.5$ & $< 8.9$  \\
 II & $[-18.2,-17.5]$ & $[8.9, 9.4]$ \\
 III& $[-19.1,-18.2]$ & $[9.4, 10.0]$ \\
 IV & $< -19.1$ & $> 10.0$ \\
\hline
\end{tabular}
\end{table}

\begin{table}
\centering
\caption{The range of 
$(g-r)_{0.1}$, $D(4000)$, $R_{90}/R_{50}$, and $\log\mu_{*}$ of
the SDSS galaxies belonging to each subsample with the constraint of
$9.4\le\log (M_{*}/M_{\odot})\le 10.0$\label{tab:sub2}}
\begin{tabular}{@{}ccccc@{}}
\hline
 Sample & $(g-r)_{0.1}$ & $D(4000)$ & $R_{90}/R_{50}$ & $\log\mu_{*}$\\ \hline
 I  & $< 0.42$ & $< 1.25$ & $< 2.18$ & $< 8.30$ \\
 II & $[0.42, 0.53]$ & $[1.25, 1.34]$ & $[2.18, 2.40]$ & $[8.30, 8.57]$\\
 III& $[0.53, 0.69]$ & $[1.34, 1.52]$ & $[2.40, 2.65]$ & $[8.57, 8.88]$\\
 IV & $> 0.69$ & $> 1.52$ & $> 2.65$ & $>
 8.88$\\ \hline
\end{tabular}
\end{table}

\section{Correlation  of  the  large-scale  tidal  shear  with  galaxy
properties}

\begin{figure}
\begin{center}
\includegraphics[width=0.5\textwidth]{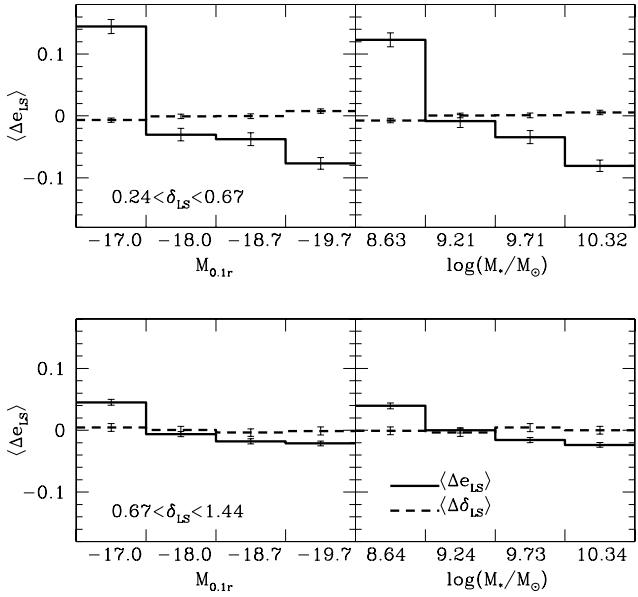}
\end{center}
\caption{Mean of the large-scale shear difference
($\Delta e_{\rm LS}\equiv e_{\rm LS}-\bar{e}_{rm LS}$) averaged over
four different samples of the SDSS galaxies (solid histogram) divided by 
their r-band absolute magnitude (left), stellar mass (right). The
errors are calculated as one standard deviation in the measurement of
the mean values. The large scale density is confined to a narrow range:  
$0.24 <\delta_{\rm LS}<0.67$ (top) and $0.67<\delta_{\rm LS}<1.44$
(bottom). In each panel, the dashed histogram represents the mean of the
large-scale density difference averaged over the four subsamples 
with the density constrained to the same range.\label{fig:shear1}}
\end{figure}

\begin{figure}
\begin{center}
\includegraphics[width=0.5\textwidth]{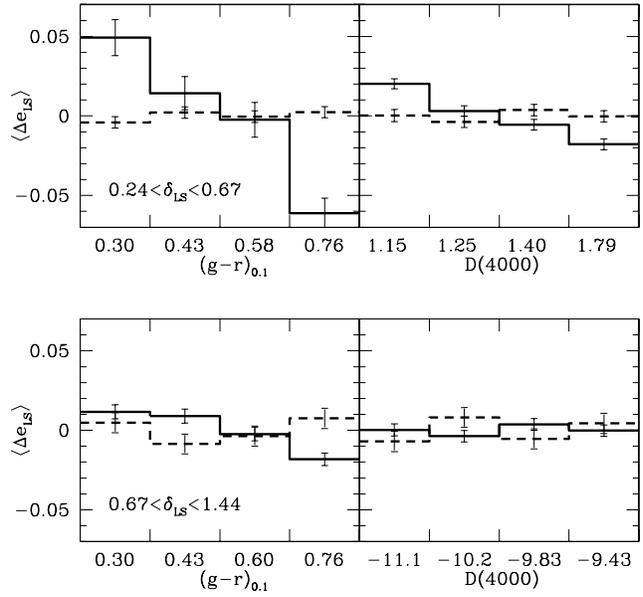}
\end{center}
\caption{Same as Fig.~\ref{fig:shear1} but the subsamples constructed 
according to the $(g-r)$ color (left) and the 4000-\AA\ break strength
(right) \label{fig:shear2}}
\end{figure}

\begin{figure}
\begin{center}
\includegraphics[width=0.5\textwidth]{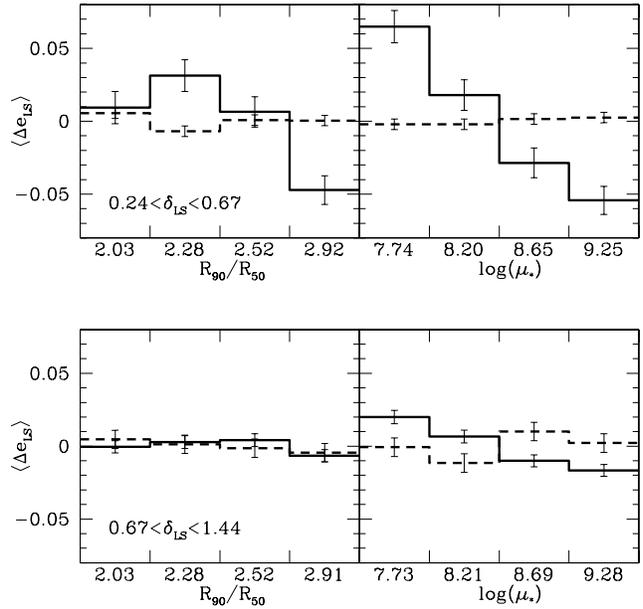}
\end{center}
\caption{Same as Fig.~\ref{fig:shear1} but the subsamples constructed 
according to the concentration parameter (left) and the surface
stellar mass density (right)\label{fig:shear3}}
\end{figure}

In  this section  we explore  the dependence  of the  large-scale {\em
tidal shear} on the same set of physical properties as in the previous
section.   In order  to normalize  out the  dependence  of large-scale
overdensity on these properties, we have divided all the galaxies into
a number of  subsamples with the large-scale overdensity  limited to a
narrow range.   Such density constrains can also  break the degeneracy
between  the trends  with galaxy  luminosity and  stellar mass  and the
trends with the other  properties. Selecting only those galaxies which
are located in the regions  whose large-scale density falls in a fixed
range, we divide these galaxies into four subsamples by binning
their galaxy properties in the same way as in \S~2,  and calculate the 
mean value  of $\Delta  e_{\rm LS}$  averaged over  each  subsample.  
Here, $\Delta  e_{\rm LS}$  is defined  as $\Delta  e_{\rm  LS}\equiv e_{\rm
LS}-\bar{e}_{\rm LS}$  where $\bar{e}_{\rm LS}$ is the  global mean of
$e_{\rm LS}$ averaged over all the selected galaxies. The results are
shown in Figs.~\ref{fig:shear1}-\ref{fig:shear3}.

Fig~\ref{fig:shear1}  plots  as solid  histograms  the  mean value  of
$\Delta  e_{\rm   LS}$  for  the  subsamples   selected  according  to
$M_{^{0.1}r}$  (left)  and  $\log(M_{*}/M_{\odot})$ (right),  and  for
overdensity constrained  to the range  of $0.24\le\delta_{LS}\le 0.67$
(top) and $0.67\le\delta\le 1.44$ (bottom). We consider only these two
ranges of $\delta_{\rm LS}$ since  in the other ranges of $\delta_{\rm
LS}  \le  0.67$ and  $\delta_{\rm  LS}\ge  1.14$,  the number  of  the
galaxies are too small to  yield reliable statistics.  As can be seen,
the   value   of  $\langle\Delta   e_{\rm   LS}\rangle$  also   varies
significantly  with  $M_{0.1r}$ and  $\log  (M_{*}/M_{\odot})$ in  the
density range of $0.24\le \delta_{\rm LS}\le 0.67$. In the other range
of $0.67\le \delta_{\rm LS}\le  1.14$, the variation of $\langle\Delta
e_{\rm  LS}\rangle$ is  weaker  but still  exists  to a  nonnegligible
degree, that is, more  luminous galaxies are preferentially located in
the  regions with  lower ellipticity.  To examine  whether or  not the
effect of the  large-scale density has been removed  from the signals,
the value of $\langle\Delta\delta_{\rm  LS}\rangle$ is also plotted as
dashed  histogram  in each  panel.   As can  be  seen,  the values  of
$\langle\Delta\delta_{\rm  LS}\rangle$  are  close  to  zero  for  all
luminosities  and all  stellar  masses, indicating  that the  observed
variations   of   $\langle\Delta e_{\rm   LS}\rangle$   between the
subsamples are not due to the effect of the large-scale density.

Fig~\ref{fig:shear2} plots the  same thing as Fig~\ref{fig:shear1} but
the four subsamples are selected to $(g-r)_{0.1}$ (left) and $D(4000)$
(right).   As  can  be   seen,  the  value  of  $\langle\Delta  e_{\rm
LS}\rangle  $ also  varies significantly  with $(g-r)_{0.1}$  when the
density range  is limited  to a narrow  range of  $0.24\le \delta_{\rm
LS}\le  0.67$. The  dependence on  $D(4000)$  is weaker  than that  on
clour, but  is still significant.  The galaxies with redder  colour or
higher  value   of  $D(4000)$  are  preferentially   located  in  the
large-scale  environment  with  lower  ellipticities.  In  the  higher
density range  of $0.67\le \delta_{\rm LS}\le 1.14$,  the variation of
$\langle\Delta e_{\rm LS}\rangle$ is quite weak, almost negligible.

Fig~\ref{fig:shear3}  shows  the   results  for  the  four  subsamples
selected according  to concentration parameter  $R_{90}/R_{50}$ (left)
and  surface stellar mass  density $\log\mu_{*}$  (right).  As  can be
seen,  the value  of $\langle\Delta  e_{\rm LS}\rangle  $  also varies
significantly with  $\log\mu_{*}$ when the  density range is  given as
$0.24\le   \delta_{\rm  LS}\le  0.67$,   whereas  the   dependence  on
$R_{90}/R_{50}$ is  much weaker. This indicates that  the surface mass
density  is  more  susceptible   to  the  effect  of  the  large-scale
shear.  The  galaxies  with  high  surface stellar  mass  density  are
preferentially  located  in   the  large-scale  environment  with  low
ellipticities.  In the  higher density  range of  $0.67\le \delta_{\rm
LS}\le 1.14$,  the variation  of $\langle\Delta e_{\rm  LS}\rangle$ is
almost negligible.

\section{Variation with the Large Scale Structure}

Another  interesting  issue  to  address  is  the  dependence  of  the
properties of galaxies on the  type of the large scale structure (LSS)
where the galaxies are found.To  classify the type of LSS, we consider
the  signs  of  the  three eigenvalues,  $\lambda_{1},\  \lambda_{2},\
\lambda_{3}$  of the tidal  shear tensor  at the  location of  a given
galaxy. We  use the  following criteria to  classify the  given region
into a void, a sheet, a filament and a halo.
\begin{eqnarray}
{\rm void}\quad &&{\rm if} \quad \lambda_{1} < 0, \\
{\rm sheet}\quad &&{\rm if} \quad \lambda_{1} > 0 \quad \& \quad \lambda_{2} < 0, \\
{\rm filament}\quad &&{\rm if} \quad \lambda_{2} > 0 \quad \& \quad \lambda_{3} < 0, \\
{\rm halo} \quad &&{\rm if} \quad \lambda_{3} > 0 
\end{eqnarray}
Table~\ref{tab:vsfh} lists  the mean value of  the large-scale density
($\bar{\delta}_{\rm  LS}$)  and  the  total  number  of  the  galaxies
($N_{g}$) located in  the void ($V$), sheet ($S$),  filament ($F$) and
halo ($H$).

\begin{figure}
\begin{center}
\includegraphics[width=0.5\textwidth]{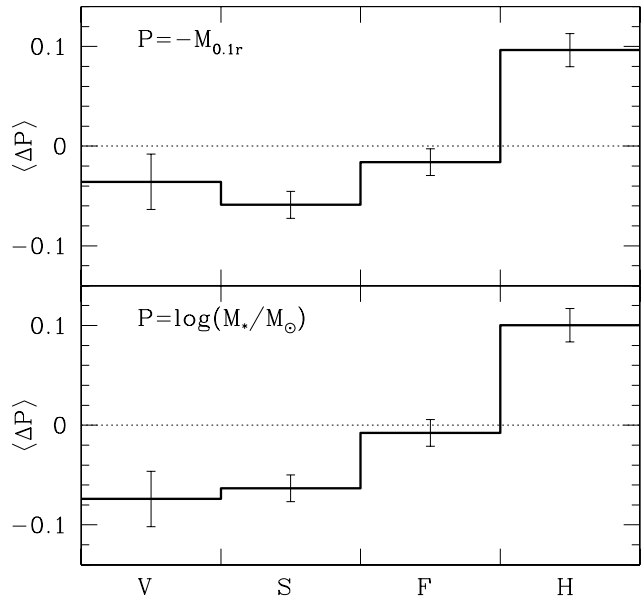}
\end{center}
\caption{Mean  values  of  the  r-band absolute  magnitude  (top)  and
stellar mass (bottom)  of the SDSS galaxies located  in the voids (V),
the sheets (S),  the filaments (F) and the halos  (H), relative to the
global mean values averaged over the whole galaxy sample.
\label{fig:vsfh1}}
\end{figure}

In Fig.~\ref{fig:vsfh1}, we show  the relative mean values of $r$-band
absolute magnitude  (top) and stellar  mass (bottom) for  the galaxies
located  in  voids, sheets,  filaments  and  haloes separately.  Here,
$\Delta  P$  is  defined  as  $\Delta  P\equiv  P-\bar{P}$  where  $P$
represents the  value of the  physical quantity and $\bar{P}$  is the
mean of  $P$ averaged over the  whole galaxy sample.  We  see that the
most luminous  (massive) galaxies  are found in  halo-like structures,
whereas the least luminous galaxies are located in voids.

\begin{table}
\centering
\caption{The LSS type, the mean of the large scale density, and the total 
number of the SDSS galaxies \label{tab:vsfh}}
\begin{tabular}{@{}cccc@{}}
\hline
Sample & LSS-Type & $\bar{\delta}_{LS}$ & $N_{\rm g}$\\ \hline
 V & Void & $-0.846\pm 0.005$  & $1919$ \\
 S & Sheet & $-0.395\pm 0.004$ & $8797$\\
 F & Filament& $0.457\pm 0.005$ & $9892$ \\
 H & Halo & $1.800\pm 0.013$ & $7746$ \\ 
\hline
\end{tabular}
\end{table}

\begin{figure}
\begin{center}
\includegraphics[width=0.5\textwidth]{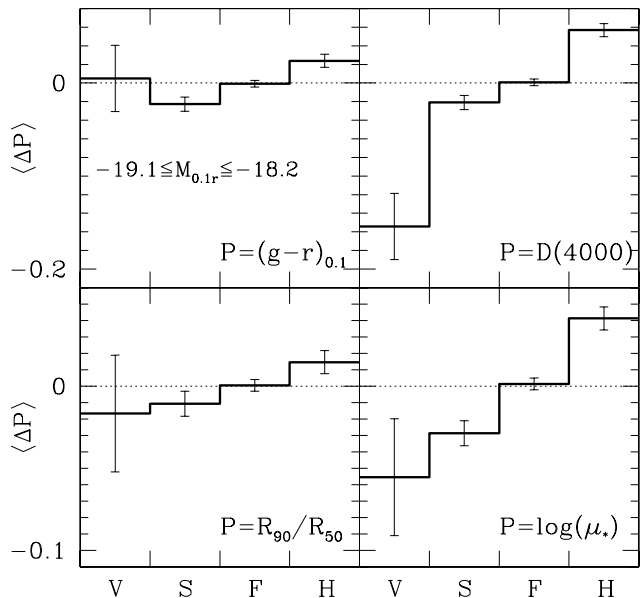}
\end{center}
\caption{Relative mean values of the $(g-r)$ color (top-left), 4000-\AA\ break
strength (top-right), concentration parameter (bottom-left), 
surface stellar mass density (bottom-right) of the SDSS galaxies 
with absolute r-band magnitude in the range of $[-19.1,-18.2]$, 
located in the voids (V), the sheets (S), the filaments (F) 
and the halos (H). \label{fig:vsfh2}}
\end{figure}

Fig.~\ref{fig:vsfh2} shows how the  average of the other four physical
parameters, $(g-r)_{0.1}$, $D(4000)$, $R_{90}/R_{50}$ and $\mu_\ast$,
relative to their  global means averaged over the  whole galaxy sample
changes among different types of the surrounding LSS.  The analysis is
performed  only for  the third  luminosity subsample  (Sample  III) in
Table 1 to remove the  effect of density-galaxy property relations. We
see that the  variation of $D(4000)$ and $\mu_\ast$  with the type of
LSS is quite significant. The mean values of $D(4000)$ and $\mu_\ast$
are highest in  halos and lowest in voids. Whereas  the mean values of
$(g-r)_{0.1}$ and  $R_{90}/R_{50}$ depend only weakly on  the types of
LSS. It  suggests that the  quantities, $D(4000)$ and  $\mu_\ast$ are
better indicators of the relation with the types of LSS.

\section{Summary and Conclusion}

By using a sample of low-$z$ galaxies drawn from the Sloan Digital Sky
Survey  and the large-scale  tidal field  reconstructed in  real space
from  the 2Mass  Redshift Survey,  we have  studied  the relationships
between  the large-scale  environment of  galaxies and  their physical
properties.  The large-scale environment  is expressed in terms of the
overdensity,  the  ellipticity  of  the  shear and  the  type  of  the
large-scale structure.  The  physical properties analyzed here include
the $r$-band  absolute magnitude, the stellar  mass, the concentration
parameter and the surface stellar mass density.

Our results can be summarized as follows.
\begin{itemize}

  \item Both luminosity and stellar mass are found to be statistically
linked  to   the  large-scale  environment,  regardless   of  how  the
environment  is quantified.  More luminous  (massive)  galaxies reside
preferentially   in   the  regions   with   higher  densities,   lower
ellipticities and halo-like structures.  

  \item  At  fixed  luminosity,  the large-scale  overdensity  depends
strongly on  parameters related to the recent  star formation history,
that is colour and 4000-\AA\  break strength, but is almost independent
of the  structural parameters, concentration and  surface stellar mass
density.  This is well  consistent with the findings of \citet{Li-06}.

\item All  the physical  properties considered here  are statistically
linked  to the  shear of  the  large-scale environment  even when  the
large-scale density  is constrained to a narrow  range.  Galaxies with
red colours,  high values  of $D(4000)$, concentrated  morphology and
high   surface  stellar   mass   density  tend   to   be  located   in
low-ellipticity regions.   This statistical link has been  found to be
most  significant in  the quasi-linear  regions where  the large-scale
density is approximately an order of unity: $\delta_{\rm LS}\sim 1$.In
highly nonlinear-regime with  $\delta_{\rm LS}\gg 1$, the significance
of this link diminishes.

\item The  properties of the galaxies  are also found to  be linked to
the types of  the large-scale structure where they  are embedded, even
when  the  r-band  absolute  magnitude  is  constrained  to  a  narrow
range. The halo galaxies tend to have highest mean values of $D(4000)$
and surface stellar mass density,  while the void galaxies have lowest
mean values  of $D(4000)$  and surface stellar  mass density.  For the
filament  galaxies, the mean  values are  found to  be similar  to the
global mean averaged over the whole sample.
\end{itemize}

Our  results have suggested  that the  galaxy-environment correlations
are induced not only by the evolutionary processes like galaxy merging
and   interactions,    but   also   by    the   initial   cosmological
conditions. Since galaxies  are distributed on the largest  scale in a
tidally induced filamentary cosmic web, the large-scale tidal field is
likely to affect  the galaxy properties and link them  to the shear of
large-scale  environment.  This  initially  induced galaxy-environment
link,  however,  is  apt  to  be overwhelmed  and  superseded  by  the
small-scale nonlinear processes especially in rich environments.

This  work has  had to  be limited  to low  redshift  galaxies ($z\leq
0.04$),  because of  a similar  redshift limit  of the  2Mass Redshift
Survey  from which  the  real-space tidal  field  was constructed.  To
understand the true role of the large-scale shears in establishing the
variations of  galaxy properties, it will be  necessary to reconstruct
the tidal field  at higher redshifts and investigate  the evolution of
the  variations   of  the  galaxy  properties   with  the  large-scale
shears. Another limitation of the current work is that, because of the
small size  of the sample,  the subsamples at ``fixed''  luminosity or
overdensity  still cover  a  relatively wide  range  of luminosity  or
overdensity, and so the correlation of luminosity with overdensity can
not  be   completely  removed   when  analyzing  the   other  physical
properties.  We  plan to reconstruct the real-space  density and shear
fields  with the  SDSS data,  and hopefully  we will  be able  to make
significant progress wit the new data.

A final conclusion is that our  results may provide a new insight into
the  galaxy formation  in a  cosmic web,  which is  still  shrouded in
mystery.
 
\section*{Acknowledgments}
J.L. thank P. Erdogdu very  much for providing the 2MRS density field.
J.L. is also very grateful to the warm hospitality of S.D.M. White and
the Max Planck Institute for  Astrophysics (MPA) in Garching where this
work was initiated  and performed.  J.L. is supported  by the visiting
scientist fellowship  from MPA in  Garching and the Korea  Science and
Engineering Foundation  (KOSEF) grant funded by  the Korean Government
(MOST,  NO.  R01-2007-000-10246-0).   C.L. is  supported by  the Joint
Postdoctoral  Programme  in  Astrophysical  Cosmology  of  Max  Planck
Institute for Astrophysics  and Shanghai Astronomical Observatory, and
by   NSFC    (10533030,   10643005,   10633020)    and   973   Program
(No.2007CB815402).

Funding for  the SDSS and SDSS-II  has been provided by  the Alfred P.
Sloan Foundation, the Participating Institutions, the National Science
Foundation, the  U.S.  Department of Energy,  the National Aeronautics
and Space Administration, the  Japanese Monbukagakusho, the Max Planck
Society, and  the Higher Education  Funding Council for  England.  The
SDSS Web  Site is  http://www.sdss.org/.  The SDSS  is managed  by the
Astrophysical    Research    Consortium    for    the    Participating
Institutions. The  Participating Institutions are  the American Museum
of  Natural History,  Astrophysical Institute  Potsdam,  University of
Basel,   Cambridge  University,   Case  Western   Reserve  University,
University of Chicago, Drexel  University, Fermilab, the Institute for
Advanced   Study,  the  Japan   Participation  Group,   Johns  Hopkins
University, the  Joint Institute  for Nuclear Astrophysics,  the Kavli
Institute  for   Particle  Astrophysics  and   Cosmology,  the  Korean
Scientist Group, the Chinese  Academy of Sciences (LAMOST), Los Alamos
National  Laboratory, the  Max-Planck-Institute for  Astronomy (MPIA),
the  Max-Planck-Institute  for Astrophysics  (MPA),  New Mexico  State
University,   Ohio  State   University,   University  of   Pittsburgh,
University  of  Portsmouth, Princeton  University,  the United  States
Naval Observatory, and the University of Washington.

\bsp

\label{lastpage}

\end{document}